\documentclass{JHEP3}

\def \be {\begin{equation}}
\def \ee {\end{equation}}
\def \bea {\begin{eqnarray}}
\def \eea {\end{eqnarray}}
\def \nn {\nonumber}

\def \a {\alpha}
\def \b {\beta}

\def \d {\delta}

\def \m {\mu}
\def \n {\nu}
\def \k {\kappa}

\def \s {\sigma}
\def \r {\rho}
\def \o {\omega}

\def \th {\theta}
\def \Th {\Theta}

\def \t {\tau}
\def \dag {\dagger}
\def \p {\partial}

\def\bd{\begin{document}}
\def\ed{\end{document}}
\def\nn{\nonumber}
\def\bea{\begin{eqnarray}}
\def\eea{\end{eqnarray}}
\let\bm=\bibitem
\let\la=\label

\def\N{{\cal N}}
\def\sst{\scriptscriptstyle}
\def\thetabar{\bar\theta}
\def\Tr{{\rm Tr}}
\def\one{\mbox{1 \kern-.59em {\rm l}}}

\def\a{\alpha}      \def\da{{\dot\alpha}}
\def\b{\beta}       \def\db{{\dot\beta}}
\def\c{\gamma}  \def\C{\Gamma}  \def\cdt{\dot\gamma}
\def\d{\delta}  \def\D{\Delta}  \def\ddt{\dot\delta}
\def\e{\epsilon}        \def\vare{\varepsilon}
\def\f{\phi}    \def\F{\Phi}    \def\vvf{\f}
\def\h{\eta}
\def\k{\kappa}
\def\L{\Lambda}
\def\m{\mu} \def\n{\nu}
\def\o{\omega}
\def\P{\Pi}
\def\r{\rho}
\def\s{\sigma}  \def\S{\Sigma}
\def\t{\tau}
\def\th{\theta} \def\Th{\Theta} \def\vth{\vartheta}
\def\X{\Xeta}
\def\z{\zeta}
\def\w{\wedge}
\def\u{\underline}
\def\hs{\hspace}

\def\cA{{\cal A}} \def\cB{{\cal B}} \def\cC{{\cal C}}
\def\cD{{\cal D}} \def\cE{{\cal E}} \def\cF{{\cal F}}
\def\cG{{\cal G}} \def\cH{{\cal H}} \def\cI{{\cal I}}
\def\cJ{{\cal J}} \def\cK{{\cal K}} \def\cL{{\cal L}}
\def\cM{{\cal M}} \def\cN{{\cal N}} \def\cO{{\cal O}}
\def\cP{{\cal P}} \def\cQ{{\cal Q}} \def\cR{{\cal R}}
\def\cS{{\cal S}} \def\cT{{\cal T}} \def\cU{{\cal U}}
\def\cV{{\cal V}} \def\cW{{\cal W}} \def\cX{{\cal X}}
\def\cY{{\cal Y}} \def\cZ{{\cal Z}}

\def\ua{\underline{\alpha}} \def\ubb{\underline{\beta}}
\def\ug{\underline{\gamma}}
\def\ub{\underline{\phantom{\alpha}}\!\!\!\beta}
\def\uc{\underline{\phantom{\alpha}}\!\!\!\gamma}
\def\um{\underline{\mu}} \def\un{\underline{\nu}}
\def\ud{\underline\delta}
\def\ue{\underline\epsilon}
\def\una{\underline a}\def\unA{\underline A}
\def\unb{\underline b}\def\unB{\underline B}
\def\unc{\underline c}\def\unC{\underline C}
\def\und{\underline d}\def\unD{\underline D}
\def\une{\underline e}\def\unE{\underline E}
\def\unf{\underline{\phantom{e}}\!\!\!\! f}\def\unF{\underline F}
\def\unm{\underline m}\def\unM{\underline M}
\def\unn{\underline n}\def\unN{\underline N}
\def\unp{\underline{\phantom{a}}\!\!\! p}\def\unP{\underline P}
\def\unq{\underline{\phantom{a}}\!\!\! q}
\def\unQ{\underline{\phantom{A}}\!\!\!\! Q}
\def\unH{\underline{H}}
\def\ul{\underline}

\def\As {{A \hspace{-6.4pt} \slash}\;}
\def\bs {{b \hspace{-6.4pt} \slash}\;}
\def\Ds {{D \hspace{-6.4pt} \slash}\;}
\def\ds {{\del \hspace{-6.4pt} \slash}\;}
\def\ss {{\s \hspace{-6.4pt} \slash}\;}
\def\ks {{ k \hspace{-6.4pt} \slash}\;}
\def\ps {{p \hspace{-6.4pt} \slash}\;}
\def\pas {{{p_1} \hspace{-6.4pt} \slash}\;}
\def\pbs {{{p_2} \hspace{-6.4pt} \slash}\;}

\def\Fh{\hat{F}}
\def\Vh{\hat{V}}
\def\Xh{\hat{X}}
\def\ah{\hat{a}}
\def\xh{\hat{x}}
\def\yh{\hat{y}}
\def\ph{\hat{p}}
\def\xih{\hat{\xi}}

\def\psit{\tilde{\psi}}
\def\Psit{\tilde{\Psi}}
\def\tht{\tilde{\th}}

\def\At{\tilde{A}}
\def\Qt{\tilde{Q}}
\def\Rt{\tilde{R}}
\def\Nt{\tilde{N}}

\def\at{\tilde{a}}
\def\st{\tilde{s}}
\def\ft{\tilde{f}}
\def\pt{\tilde{p}}
\def\qt{\tilde{q}}
\def\vt{\tilde{v}}
\def\nt{\tilde{n}}

\def\delb{\bar{\partial}}
\def\bz{\bar{z}}
\def\bD{\bar{D}}
\def\bB{\bar{B}}

\def\bk{{\bf k}}
\def\bl{{\bf l}}
\def\bp{{\bf p}}
\def\bq{{\bf q}}
\def\br{{\bf r}}
\def\bx{{\bf x}}
\def\by{{\bf y}}
\def\bR{{\bf R}}
\def\bV{{\bf V}}

\def\d{\delta}\def\D{\Delta}\def\ddt{\dot\delta}

\def\p{\partial} \def\del{\partial}
\def\xx{\times}
\def\uno{\mbox{1 \kern-.59em {\rm l}}}

\def\trp{^{\top}}
\def\inv{^{-1}}
\def\dag{{^{\dagger}}}
\def\pr{\prime}

\def\rar{\rightarrow}
\def\lar{\leftarrow}
\def\lrar{\leftrightarrow}
\title{Semi-classical strings in $AdS_4\times CP^3$}
\author{Bin Chen\\
Department of Physics,\\
and State Key Laboratory of Nuclear Physics and Technology,\\
Peking University,\\
Beijing 100871, P.R. China\\
\email{bchen01@pku.edu.cn}}

\author{Jun-Bao Wu\\International School for Advanced Studies (SISSA), \\
via Beirut 2-4, I-34014 Trieste, Italy\\
and INFN, Trieste section, Via Valerio 2, I - 34127 Trieste, Italy\\
\email{wujunbao@sissa.it}}

\date{\today}

\abstract{ In this paper, we study the semi-classical strings in
$AdS_4\times CP^3$ spacetime. We construct various kinds of string
solutions, including the point-like, circular, folded and pulsating
strings. For the circular and folded strings, we figure out their
field theory dual operators. In particular,  we discuss the
anomalous dimensions of the corresponding operators from decoupled
$SU(2)\times SU(2)$ spin chain. We find that in the large angular
momentum limit, the field theory and string theory results are in
very good agreement up to an interpolating function of coupling
constant. }

\preprint{\ SISSA-43/2008/EP}
\newpage

\begin{document}
\section{Introduction}

The semi-classical strings has played an important role in studying
$AdS_5/SYM_4$ correspondence. In a remarkable paper \cite{GKP}, it
was found that the fluctuations around the point-like string
captured the right string spectrum of BMN plane-wave background
\cite{BMN}. Moreover, the study of the multi-spin string solutions
reveals that even when the configuration is far from BPS, the energy
of the string solution with large angular momentum could be in
perfect agreement with the ones calculated in dual gauge theory. The
agreement beyond BPS limit relies on the fact that on string side
the quantum corrections of strings was suppressed by the large
quantum number, while on the field theory side the dual field
operators to these string solutions are the composite operators,
whose anomalous dimension matrix (ADM) could be related to the
Hamiltonian of integrable spin chain. Moreover this agreement
suggest that  both sides of $AdS_4/SYM_4$ correspondence are
integrable. There are lots of study on this topic. Please see
\cite{Tseytlin} for a nice review and references.

Very recently, inspired by the study of Bagger-Lambert-Gustavson
theory on $N$ membranes \cite{BL,G,3algebra}, in \cite{ABJM} Aharony
et.al. proposed a ${\cal N}=6$ Chern-Simons theory coupled with
bi-fundamental matter, describing $N$ membrane on $S^7/Z_k$. In
particular, it was pointed out that in the limit with $k\ll N\ll
k^5$, this theory is dual to IIA string theory in $AdS_4\times
CP^3$. Many aspects of the BPS sector of this new $AdS_4/CFT_3$
correspondence have been studied in \cite{ABJM}. Quite recently, the
near-BPS sector are also studied by using the Penrose limit of this
IIA string theory background \cite{Nishioka:2008gz, Gaiotto:2008cg, Grignani:2008is}.
Other relevant work could be found in
\cite{Benna:2008zy}-\cite{Ahn:2008gd}. In particular, in
\cite{Minahan:2008hf, Gaiotto:2008cg} it was pointed out there exist
integrable spin-chain structure in the field theory. It would be
interesting to go beyond the near BPS sector and investigate the
semiclassical configurations and their implications in $AdS_4/CFT_3$
correspondence.

In this paper, we will study the semi-classical string solutions in
$AdS_4\times CP^3$. We will mainly focus on the spinning solutions
in $CP^3$. We will calculate their energy and angular momenta and
compare with their field theory duals. This is possible due to the
recent study of 2-loop integrable structure in the field
theory\cite{Minahan:2008hf, Gaiotto:2008cg}. Our solutions include a
point-like one which is BPS and corresponding to the ground states
of the string theory in the IIA plan wave background obtained by
taking  Penrose limit. We also study the circular and folded string
configurations. We find that some of our solutions are far from BPS,
but still has a field theory dual. For the circular string and
folded string, we discuss the energy of the corresponding operator
from algebraic Bethe Ansatz equation(ABAE). The agreement between
string theory and field theory results could be perfect if an
interpolating function of coupling constant is introduced.

The paper will be organized as follows. In section 2, we will set up
our system and have a general discussion of the semi-classical
solutions. In section 3, we will present several kinds of spinning
solutions. In section 4, we will discuss the field dual to some of
these spinnning string solutions. We will end with conclusion and discussion in
section 5.

\section{Action and equations of motion}

The three-dimensional ${\cal N}=6$ superconformal theory proposed by
ABJM in \cite{ABJM} is a Chern-Simons theory with gauge group
$SU(N)\times SU(N)$ with bifundamental superfields $A_1, A_2$ and
anti-bifundamental superfields $B_1, B_2$. The action has a pure
Chern-Simons part:
 \bea
 S_{CS}&=&\frac{k}{4\pi}\int (A_{(1)}\wedge dA_{(1)}+\frac23 A_{(1)}^3
-A_{(2)}\wedge
 dA_{(2)}- \frac23 A_{(2)}^3),
 \eea
 and the superpotential
 \bea
 W&=&\frac{4\pi}{k}\Tr(A_1B_1A_2B_2-A_1B_2A_2B_1).
 \eea
 Note that the sign before the Chern-Simons terms are opposite and
 the superpotential has actually an $SU(2)\times SU(2)$ global
 symmetry, which acting on the $A$'s and the $B$'s separately.
 The level of the Chern-Simons
theory for the two components of the gauge group are $k$ and $-k$,
respectively. When $k$ and $N$ satisfy $k\ll N\ll k^5$, this field
theory is dual to IIA superstring theory on $AdS_4\times CP^3$ with
constant dilaton, RR two-form and four-form fluxes. The constant
dilaton reads: \be e^{2\phi}=2^{5/2}\pi N^{1/2}k^{-5/2}. \ee Since
we are interested in semiclassical fundamental string solution in
this background, the RR fluxes play no roles here.

Let us start from the metric of $AdS_4\times CP^3$,
\bea ds^2&=&\frac14 R^2(- \cosh^2\rho dt^2+d\rho^2+\sinh^2\rho(d\theta^2+\sin^2\theta d\phi^2))+R^2(d\xi^2+\nn\\
&&\cos^2\xi\sin^2\xi
(d\psi+\frac12\cos\theta_1d\varphi_1-\frac12\cos\theta_2d\varphi_2)^2+\frac14\cos^2\xi(d\theta_1^2+\sin^2\theta_1
d\varphi_1^2)+\nn\\
&&\frac14\sin^2\xi(d\theta_2^2+\sin^2\theta_2 d\varphi_2^2)). \eea
The radius $R$ here is\footnote{We take $\alpha^\prime=1$.} \be
R=2^{5/4}\pi^{1/2}\lambda^{1/4}, \ee where $\lambda\equiv N/k$ is
the 't Hooft coupling constant of this three dimensional
superconformal theory. In the above metric, the $CP^3$ part is the
standard Fubini-Study metric which could be obtained in the
following way. We first consider a unit $S^7$ \be
\sum_{i=1}^4|Z_i|^2=1,\ee embedded in $C^4$. Then we parameterize
the complex coordinates $Z_i$'s as follows: \bea
Z_1&=&\cos\xi\cos\frac{\theta_1}2\exp[i(y+\frac{\psi+\varphi_1}2)],\label{z1}\\
Z_2&=&\cos\xi\sin\frac{\theta_1}2\exp[i(y+\frac{\psi-\varphi_1}2)],\\
Z_3&=&\sin\xi\cos\frac{\theta_2}2\exp[i(y+\frac{-\psi+\varphi_2}2)],\\
Z_4&=&\sin\xi\sin\frac{\theta_2}2\exp[i(y+\frac{-\psi-\varphi_2}2)].\label{z4}
\eea Here $0\le\xi<\frac\pi2, 0\le y<2\pi, -2\pi<\psi<2\pi$ and
$(\theta_i, \varphi_i)$ are coordinates of two $S^2$'s. Now the
induced metric on $S^7$ can be written as a $U(1)$ fiber over
$CP^3$: \be ds^2_{S^7}=ds^2_{CP^3}+(dy+A)^2. \ee In this way, we get
the metric on $CP^3$ as above (here $A$ is a one-form).

For most of our solutions, we assume that $t, \phi, \psi, \varphi_1,
\varphi_2$ are function of $\tau$ only and $\rho, \xi, \theta_1,
\theta_2$ are only the periodic functions of $\sigma$. The bosonic
part of the string Lagrangian is \be
L_B=\frac{1}{2}\sqrt{-g}g^{ab}G_{MN}\p_aX^M\p_bX^N \ee where
$G_{MN}$ is the background metric above and $g_{ab}$ is the
worldsheet metric. We can choose the conformal gauge, in which
$g_{ab}=e^\gamma Diag(-1,1)$. Then we have the action:
\bea S&=&\frac{\sqrt{\tilde{\lambda}}}{4\pi}\int d\sigma d\tau(\frac{1}{4}\cosh^2\rho \dot{t}^2+\frac14\rho^{\prime 2}+\frac14\sinh^2\rho(\theta^{\prime 2}-\sin^2\theta\dot{\phi}^2)+\xi^{\prime 2}\nn\\
&&-\cos^2\xi\sin^2\xi(\dot{\psi}+\frac12\cos\theta_1\dot{\varphi_1}-\frac12\cos\theta_2\dot{\varphi_2})^2+\frac14\cos^2\xi(\theta^{\prime 2}_1-\sin^2\theta_1\dot{\varphi}^2_1)+\nn\\
& &\frac14\sin^2\xi(\theta^{\prime
2}_2-\sin^2\theta_2\dot{\varphi}^2_2)). \eea The relation between
 $\tilde{\lambda}$ and the 't
Hooft coupling $\lambda$ is: \be \tilde{\lambda}^2=32\pi^2\lambda^2.
\ee

From the metric, we know that the background has at least five
Killing vectors corresponding to the translations along
$t,\phi,\psi,\varphi_1,\varphi_2$. The Killing vector along $t$
gives the conserved energy and the other four Killing vectors give
the conserved angular momenta. Let us make ansatz:
\begin{equation}
t=\kappa\tau,\hspace{3ex}\phi=v\tau,\hspace{3ex}\psi=\omega_1\tau,\hspace{3ex}\varphi_1=\omega_2\tau,\hspace{3ex}\varphi_2=\omega_3\tau
\end{equation}
With this setup, we have the conserved quantities:
\begin{eqnarray}
E&=&\frac14\cosh^2\rho\sqrt{\tilde{\lambda}}\kappa\\
S&=&\sqrt{\tilde{\lambda}}\int \frac{d\sigma}{2\pi}v\sinh^2\rho\sin^2\theta\\
J_1&=&\sqrt{\tilde{\lambda}}\int \frac{d\sigma}{2\pi}\cos^2\xi\sin^2\xi(\omega_1+\frac{\cos\theta_1}{2}\omega_2-\frac{\cos\theta_2}{2}\omega_3)\\
J_2&=&\sqrt{\tilde{\lambda}}\int
\frac{d\sigma}{2\pi}[\frac{1}{4}\cos^2\xi\sin^2\theta_1\omega_2+\cos^2\xi\sin^2\xi(\frac{\cos^2\theta_1}{4}\omega_2
+\frac{\cos\theta_1}{2}(\omega_1-\frac{\cos\theta_2}{2}\omega_3))]\nonumber\\
J_3&=&\sqrt{\tilde{\lambda}}\int
\frac{d\sigma}{2\pi}[\frac{1}{4}\sin^2\xi\sin^2\theta_2\omega_3+\cos^2\xi\sin^2\xi(\frac{\cos^2\theta_2}{4}\omega_3
-\frac{\cos\theta_2}{2}(\omega_1+\frac{\cos\theta_1}{2}\omega_2))]\nonumber
\end{eqnarray}
The equations of motion for $t, \phi, \psi, \varphi_1, \varphi_2$ are just the conservation of $E, S, J_i$'s.

Now we turn to the equations of motion for $\rho, \theta,
\xi,\theta_1, \theta_2$. The equations of motion are:
\bea\rho^{\prime\prime}&=&\sinh\rho\cosh\rho
\dot{t}^2-\sinh\rho\cosh\rho\sin^2\theta\dot{\phi}^2+\sinh\rho\cosh\rho
\theta^{\prime 2},\\
 \frac{\partial}{\partial\sigma}(\frac12\sinh^2\rho\theta^\prime)&=&
\frac12\sinh^2\rho\sin\theta\cos\theta, \\
2\xi^{\prime\prime}&=&-\frac12\sin4\xi(\dot{\psi}+\frac12\cos\theta_1\dot{\varphi}_1-\frac12\cos\theta_2\dot{\varphi}_2)^2
+\frac12\cos\xi\sin\xi\sin^2\theta_1\dot{\varphi}_1^2\nn\\&-&\frac12\sin\xi\cos\xi\sin^2\theta_2\dot{\varphi}_2^2
-
\frac12\sin\xi\cos\xi\theta^{\prime 2}_1+\frac12\sin\xi\cos\xi\theta^{\prime 2}_2, \\
\frac{\partial}{\partial\sigma}(\frac12\cos^2\xi\theta^{\prime}_1)&=&\cos^2\xi\sin^2\xi
\sin\theta_1(\dot{\psi}+\frac12\cos\theta_1\dot{\varphi_1}-\frac12\cos\theta_2\dot{\varphi_2})\dot{\varphi}_1\nn\\
&-&\frac12\cos^2\xi\sin\theta_1\cos\theta_1\dot{\varphi}_1^2 \\
\frac{\partial}{\partial\sigma}(\frac12\cos^2\xi\theta^{\prime}_2)&=&-\cos^2\xi\sin^2\xi
\sin\theta_2(\dot{\psi}+\frac12\cos\theta_1\dot{\varphi_1}-\frac12\cos\theta_2\dot{\varphi_2})\dot{\varphi}_2\nn\\
&-&\frac12\sin^2\xi\sin\theta_2\cos\theta_2\dot{\varphi}_2^2 \eea
The Virasoro constraint \bea
G_{MN}(\p_0X^M\p_0X^N+\p_1X^M\p_1X^N)&=&0,\nn\\
G_{MN}\p_0X^M\p_1X^N&=&0, \eea gives
\begin{eqnarray}
\cosh^2\rho\kappa^2/4&=&\frac{1}{4}(\rho^{\prime
2}+\sinh^2\rho(\theta^{\prime 2}+\sin^2\theta \dot{\phi}^2))\nn\\
& &+\xi^{\prime 2}+\cos^2\xi\sin^2\xi(\omega_1+\frac{\cos\theta_1\omega_2-\cos\theta_2\omega_3}{2})^2 \nn\\
& &+\frac{1}{4}\cos^2\xi(\theta_1^{\prime 2}+\sin^2\theta_1\omega_2^2)+\frac{1}{4}\sin^2\xi(\theta_2^{\prime 2}+\sin^2\theta_2\omega_3^2)
\end{eqnarray}

\section{Various semi-classical string solution}

In this section, we would like to discuss the semi-classical string
solutions in $AdS_4\times CP^3$, with emphasis on the spinning
string solutions in $CP^3$. For the string solutions in $AdS_4$, the
discussion is very similar to the case in $AdS_5$. The only
difference lies at the fact that there is only one spin quantum
number in $AdS_4$ while there are two spins in $AdS_5$. The rotation
string and pulsating string could be constructed easily.

For the semi-classical string in $CP^3$, the construction is different. In this case, $\rho$ is constant and $t=\kappa\tau$, $,\theta, \phi$ is constant, then
the equation of motion for $\theta$ is satisfied and from the equation of motion for $\rho$, we get
\be \cosh\rho\sinh\rho\kappa=0, \ee
which restricts $\rho=0$.

Compared to the similar construction of multi-spin solutions in
$AdS_5\times S^5$\cite{Frolov2003}, the construction of spinning
solutions in $AdS_4\times CP^3$ is more tricky and restrictive,
whose dual field operators are also not transparent. We manage to
find the dual operators for point-like string, a class of circular
string and a class of folded string, by matching the global charges.

\subsection{Point-like  solution}
 Consider
the point-like solution in which there is no dependence on $\sigma$.\footnote{After we finished our paper, a preprint \cite{Gromov:2008bz} by Gromov
and Vieria appears in arXiv. The discussions about point like
strings there has some overlap with our discussions here. }
Let $\theta_1=\theta_2=0$, then the equations of motion for
$\theta_1$ and $\theta_2$ are satisfied. And from the equation of
motion for $\xi$, we get \be
\sin4\xi(\omega_1+\frac12\omega_2-\frac12\omega_3)^2=0 \ee So for
generic $\omega_i$'s, $\xi$ can only take $\pi/4$. The Virasoro
constraints give
\begin{eqnarray}
\kappa^2/4&=&\frac14 (\omega_1+\frac{\omega_2-\omega_3}{2})^2
\end{eqnarray}
Then we get \be E=J_1=2J_2=-2J_3 \label{pp}\ee

Recall that $J_i$'s is the quantum number corresponding to the Killing vector
\begin{equation}
J_1=-i \frac{\partial}{\partial\psi}
\end{equation}
\begin{equation}
 J_2=-i\frac{\partial}{\partial \varphi_1},\hspace{3ex}  J_3=-i\frac{\partial}{\partial \varphi_2}.
\end{equation}
The essential fact is that following \cite{Nishioka:2008gz}, we
can identify $X_1, X_2, X_3, X_4$ in (\ref{z1})-(\ref{z4}) as the
scalar fields $A_1, A_2, \bar{B}_1, \bar{B}_2$. Then  we can write
down the charges of the scalar fields as
\begin{equation}
J_1(A_1)=J_1(A_2)=J_1(B_1)=J_1(B_2)=1/2
\end{equation}
\begin{equation}
J_2(A_1)=-J_2(A_2)=1/2, \hspace{3ex}J_2(B_1)=J_2(B_2)=0
\end{equation}
\begin{equation}
J_3(A_1)=J_3(A_2)=0,\hspace{3ex} J_3(B_1)=-J_3(B_2)=-1/2
\end{equation}
Then the chiral operator $\Tr(A_1B_1)^J$ has $J_1=J$, $J_2=J/2$,
$J_3=-J/2$, $E=J$, in perfect match with the relation (\ref{pp}). In
fact, as the case studied in \cite{GKP}, the point-like solution is
dual to the chiral primary operators, which is BPS and can be
identified as the ground state of the IIA string in the plane-wave
background. The identification of $\Tr(A_1B_1)^J$ as the ground
state has also been pointed out in \cite{Nishioka:2008gz}. Similar
to the IIB string in $AdS_5\times S^5$, it could be expected that
the fluctuations around this point-like solution is actually the IIA
string spectrum in the plane-wave
background\cite{GKP,FT2002,Nishioka:2008gz}.

\subsection{Folded string I}

Let us try the following ansatz: $\theta_1=\theta_2=0$, then we have
\be \xi^{\prime\prime}=-\frac14\sin4\xi\tilde{\omega}^2  \ee where
$\tilde{\omega}=\omega_1+(\omega_2-\omega_3)/2$. The Virasoro
constraint is now \be \frac{\kappa^2}{4}=\xi^{\prime
2}+\frac{\sin^22\xi}{4}\tilde{\omega}^2. \ee
Since we consider the folded string here, $\xi$ will take its
maximal value at some $\xi_0$. When $\xi=\xi_0$, we have
$\xi^{\prime}=0$ and so $\kappa^2=\sin^22\xi_0\tilde{\omega}^2$.
Then we get \be \xi^{\prime
2}=\frac{\tilde{\omega}^2}{4}(\sin^22\xi_0-\sin^22\xi), \ee which
leads to \be
2\pi=4\int_0^{\xi_0}\frac{2d\xi}{\tilde{\omega}(\sqrt{\sin^22\xi_0-\sin^22\xi})}.\label{omega}
\ee The angular momenta in this case is just \bea
J_1&=&\sqrt{\tilde{\lambda}}\int \frac{d\sigma}{2\pi}\cos^2\xi\sin^2\xi\tilde{\omega}\label{j1} \\
J_2&=&\frac{J_1}{2}\\
J_3&=&-J_2 \eea The Eqs.~(\ref{omega}) and (\ref{j1}) give us \bea
\tilde{\omega}&=&\frac2{\pi}K(q),\\
J_1&=&\frac{\sqrt{\tilde{\lambda}}}{2\pi}(K(q)-E(q)), \eea where
$q=\sin^22\xi_0$ and $E(q), K(q)$ are the elliptic integrals of
first kind and second kind, respectively. The energy $E$ is \be
E=\frac{\sqrt{\tilde{\lambda}}}{4}\kappa=\frac{\sqrt{\tilde{\lambda}}}{4}\tilde{\omega}
\sin2\xi_0=\frac{\sqrt{\tilde{\lambda}}}{2\pi}\sin2\xi_0K(q). \ee
This string configuration is a folded string. The relation between
different angular momenta suggest that the solution has only one
angular momentum. It is reminiscent of the folded string solution
spinning in $S^5$ discussed in \cite{GKP}. It is interesting to
consider the large $J$ limit, which corresponds to $\xi_0 \to
\pi/4$. In this limit, one has $E\to\infty$, $J_1\to\infty$ and
$E-J_1\sim \frac{\sqrt{\tilde{\lambda}}}{2\pi}$. This looks similar
to the relation in the giant magnon case with $p=1$ in
\cite{Gaiotto:2008cg}, but actually folded string solution is very
different from the semiclassical string of giant magnon explicitly
constructed in \cite{Grignani:2008is,Grignani:2008te}.

\subsection{Circular strings}

In this subsection, we will fix $\xi=\xi_0$ being a constant. First
let us further assume that $\omega_3=0$, $\theta_1=0$, the equation
of motion for $\theta_1$ is satisfied. From the equation of motion
for $\theta_2$ we have \be\theta^{\prime\prime}_2=0,\ee so
$\theta_2=n\sigma$, (we choose $n\ne0$). The equation of motion for
$\xi$ gives: \be \cos2\xi_0=-\frac{n^2}{(2\omega_1+\omega_2)^2}. \ee
For this solution, we have: \be
J_1=\sqrt{\tilde{\lambda}}\cos^2\xi_0\sin^2\xi_0(\omega_1+\frac12\omega_2)=2J_2,
J_3=0, \ee and \be
E^2=\frac{J_1^2}{4\sin^2\xi_0\cos^2\xi_0}+\frac{\tilde{\lambda}\sin^2\xi_0n^2}{16}.\ee
This is  a string solution with one independent angular momentum.
For $n=0$, we just come back to the point-like solution discussed
before.

We can also make ansatz that $\omega_2=\omega_3=0$, then the
equations of motion for $\theta_1, \theta_2$ give
\be\theta^{\prime\prime}_1=\theta^{\prime\prime}_2=0, \ee which
gives $\theta_i=n_i\sigma$. If we choose $n_1$ and $n_2$ to be
nonzero, then $J_2=J_3=0$, The equation of motion for $\xi$ gives:
\be\cos2\xi_0=\frac{n_1^2-n_2^2}{4\omega_1^2}. \ee The relation
between $E$ and $J_1$ is \be
E^2=\frac{J_1^2}{4\sin^2\xi_0\cos^2\xi_0}+\frac{\tilde{\lambda}}{16}(\cos^2\xi_0n^2_1+\sin^2\xi_0n^2_2).\ee
This solution is a circular string with one angular momentum.
Especially if we choose $n_1=\pm n_2$, we have $\xi_0=\pi/4$ and \
\be E=\sqrt{J_1^2+\frac{\tilde{\lambda}
n_1^2}{16}}=J_1(1+\frac{\tilde{\lambda}
n_1^2}{32J_1^2}+\cdots).\label{circular}\ee In this case, the
circular string has a field theory dual. Let us consider the
composite operator in field theory $\Tr((A_1B_1)^J(A_2B_2)^J)$. It
has $J_1=2J$, $J_2=J_3=0$ and at the classical level, $E=2J$. This
is in consistent with the relations that the circular string respect
to. We will study this operator in the next section.


\subsection{Folded string II}

The study of the circular string solutions suggest that one may have
to fix $\xi=\pi/4$ in order to find their field theory dual operator
chains. In this case, from the equation of motion for $\xi$, it is
quite natural to require $\theta_1=\pm \theta_2$. Actually, this is
the only possible way to have nontrivial solution. Then from the
equations for $\theta_1$ and $\theta_2$, we find that this is only
possible for $\omega_2=-\omega_3$ and \be
\theta_1^{\prime\prime}=\sin\theta_1 \omega_1\omega_2. \ee This
equation is in consistency with the Virasoro constraint, which has
\be \kappa^2=\theta_1^{\prime
2}+\omega^2_1+\omega^2_2+2\omega_1\omega_2\cos\theta_1. \ee

Let us first consider two special case. When we take $\omega_2=0$,
this reduce to the circular string we studied before.

However, the case with $\omega_1=0$ is also interesting. In this
case, we have \bea J_1=0,
J_2=-J_3=\frac{\sqrt{\tilde{\lambda}}}{8}\omega_2,
\k^2=\omega_2^2+n^2 \eea This is another circular string solution,
quite similar to the first one, but the field theory dual is very
different. To respect the relation between quantum numbers, we are
led to considering the following operators:
$\Tr(A_1B_1)^{J_2/2}(B_2^\dagger A_2^\dagger)^{J_2/2}$. The energy
of string in the large $J_2$ limit is
 \be
 E=2J_2+\frac{n^2\tilde{\lambda}}{64J_2}+\cdots.
 \ee
  Classically the dual operator has dimension
$\Delta=2J_2$, in consistent with the zero order string energy. It
should be keep in mind that the first order correction in the string
energy is of order $\tilde{\lambda}/J_2^2$, similar to the circular
string we discussed before.

In general, the solution is a folded string configuration. Since
$\theta_1$ is periodic of $\sigma$, so we have \be -\theta_1(0)\leq
\theta_1(\sigma)\leq \theta_1(0),
\kappa^2=\omega^2_1+\omega^2_2+2\omega_1\omega_2\cos\theta_1(0). \ee
If $\omega_1\omega_2>0$, $\theta$ is varying around $\pi$. On the
contrary, if $\omega_1\omega_2<0$, the folded string is centered at
$\theta_1=0$. Without losing generality, we will assume
$\omega_1>0,\omega_2<0$ such that $\omega_1\omega_2<0$. Then, we have \bea
2\pi=\int_0^{2\pi}d\sigma=\frac{2}{\sqrt{-\omega_1\omega_2}}\int_0^{\theta_1(0)}\frac{d\theta_1}
{\sqrt{\sin^2\frac{\theta_1(0)}{2}-\sin^2\frac{\theta_1}{2}}}. \eea
This gives us \be \sqrt{-\omega_1\omega_2}=\,\frac{2}{\pi}K(x),\ee
where $x=\sin^2\frac{\theta_1(0)}{2}$.

The energy of the folded string solution is just
$E=\frac14\sqrt{\tilde{\lambda}}\kappa$. The angular momenta are
\bea
 J_1&=&\sqrt{\tilde{\lambda}}\int
 \frac{d\sigma}{2\pi}\frac14(\omega_1+\cos\theta_1\omega_2)\nn\\
 &=&\frac{\sqrt{\tilde{\lambda}}}{2\pi\sqrt{-\omega_1\omega_2}}\left((\omega_1-\omega_2)K(x)+2\omega_2E(x)
 \right),\\
J_2&=&\frac{\sqrt{\tilde{\lambda}}}{8\pi\sqrt{-\omega_1\omega_2}}\int_0^{\theta_1(0)}\frac{(\omega_2+\omega_1\cos\theta_1)d\theta_1}
{\sqrt{\sin^2\frac{\theta_1(0)}{2}-\sin^2\frac{\theta_1}{2}}} \\
&=&\frac{\sqrt{\tilde{\lambda}}}{4\sqrt{-\omega_1\omega_2}}\left((\omega_2-\omega_1)K(x)+2\omega_1E(x)\right),\\
J_3&=&-J_2. \eea
We have the following relation \be\frac{\omega_1}{2}J_1-\omega_2J_2=\frac{\sqrt{\tilde{\lambda}}}{8}(\omega_1^2-\omega_2^2)
\ee In the case that $\omega_1=-\omega_2$, we have $J_1=-2J_2$.

It is convenient to introduce the following quantities: \be
\cE=\frac{E}{\sqrt{\tilde{\lambda}}},\hspace{3ex}
\cJ_i=\frac{E}{\sqrt{\tilde{\lambda}}}, \hspace{3ex}i=1,2,3. \ee In
terms of these quantities, we have
 \bea
 \omega_1&=&K(x)\frac{K(x)\cJ_1-2\cJ_2(2E(x)-K(x))}{E(x)(K(x)-E(x))} \\
 \omega_2&=&K(x)\frac{2K(x)\cJ_2-\cJ_1(2E(x)-K(x))}{E(x)(K(x)-E(x))},
 \eea
and the following key relations
 \bea\label{relation}
 \left(\frac{\cE}{K(x)}\right)^2-\left(\frac{\cJ_1+2\cJ_2}{E(x)}\right)^2&=&\frac{4}{\pi^2}x
 \\
 \left(\frac{\cJ_1-2\cJ_2}{E(x)-K(x)}\right)^2-\left(\frac{\cJ_1+2\cJ_2}{E(x)}\right)^2&=&\frac{4}{\pi^2}.
 \eea
We will show that in the next section, due to the above relation the
folded string could be in perfect match with the dual field theory
operators up to an interpolating function.

The dual operators in the field theory is somehow subtle. To match the above
angular momenta, we propose the following identification:
\be
\left\{
\begin{array}{ll}
\Tr((A_1B_1)^{\frac{J_1}{2}+J_2}(A_2B_2)^{\frac{J_1}{2}-J_2}),&\hspace{3ex}\mbox{for $\omega_1+\omega_2>0$}\\
\Tr((B_1^\dagger
A_1^\dagger)^{-(\frac{J_1}{2}+J_2)}(A_2B_2)^{\frac{J_1}{2}-J_2}),&\hspace{3ex}\mbox{for $\omega_1+\omega_2<0$}
\end{array}
\right.
\ee
However, if we take $\omega_1=-\omega_2$, then the above operators reduce to $\Tr(A_2B_2)^{J_1}$, which is a BPS primary and has $\Delta=J_1$ without quantum correction. But from the string calculation we know that there do exist
higher order corrections.

\subsection{Pulsating string}

Before ending this section, let us discuss pulsating string, another
kind of semi-classical string solution. The pulsating string purely
in $AdS_4$ is quite similar to the one in $AdS_5$
\cite{Minahan:2002rc}. So we focus on the circular pulsating string
expanding and contracting on $CP^3$. To simplify the discussion, we
let $\theta, \phi, \theta_1, \theta_2, \varphi_1, \varphi_2$ be
fixed to zero, $t=\kappa \tau, \psi=n\sigma$ and $\rho,\xi$ be the
function of $\tau$. Then the Green-Schwarz action is \be
S=\frac{\sqrt{\tilde{\lambda}}}{4\pi}\int d\sigma
d\tau(\frac{\kappa^2}{4}\cosh^2\rho-\frac14\dot{\rho}^2-\dot{\xi}^2+\cos^2\xi\sin^2\xi
n^2) \ee which leads to the equations of motion \bea
\ddot{\rho}&=&-\kappa^2\sinh\rho\cosh\rho \\
\ddot{\xi}&=&-\frac14 n^2\sin 4\xi \eea On the other hand, the
Virasoro constraint is now \be
\frac14(-\kappa^2\cosh^2\rho+\dot{\rho}^2)+\dot{\xi}^2+\cos^2\xi\sin^2\xi
n^2=0 \ee To be consistent with the equations of motion, we notice
that $\dot{\rho}$ has to be vanishing. So we fix $\rho=0$. Then we
have only one equation to solve. Let $\eta=2\xi$, we have \be
\dot{\eta}^2+\sin^2\eta n^2-\kappa^2=0 \label{eta}\ee or \be
\ddot{\eta}=-\frac{n^2}{2}\sin 2\eta. \ee This looks nice a
one-dimensional pendulum. From eq.~(\ref{eta}), we know that when
$\eta$ take the maximal value, $\eta_0$, $\dot{\eta}=0$. Then
$\kappa^2=n^2\sin^2\eta_0$. So \be
\dot{\eta}^2=n^2(\sin^2\eta_0-\sin^2\eta). \ee If the period of this
pendulum is $T$ (this is measure by the worldsheet time $\tau$), we
have, \be
T=4\int_0^{\eta_0}\frac{d\eta}{n\sqrt{\sin^2\eta_0-\sin^2\eta}}.\ee
When $\tau\in[0, \frac\pi4]$, we have \be
\tau=\int_0^{\eta}\frac{d\tilde{\eta}}{n\sqrt{\sin^2\eta_0-\sin^2\tilde{\eta}}}.\ee

\section{Field theory dual operators}

The integrable spin chain in ABJM theory has been discussed in
\cite{Minahan:2008hf, Gaiotto:2008cg}. It was pointed out in
\cite{Minahan:2008hf}, at two-loop order, the anomalous dimension
matrix (ADM) of the composite operators constructed from the
scalar fields could be identified with an integrable Hamiltonian
of an $SU(4)$ spin chain. Including the fermions in the operators,
the spin chain is extended to $SU(2|2)$ in \cite{Gaiotto:2008cg}.
More precisely, following the notation in \cite{Minahan:2008hf}
let us consider the following gauge invariant operators of the
form
 \be
 \Tr(Y^{A_1}Y^\dagger_{B_1}Y^{A_2}Y^\dagger_{B_2}\cdots
 Y^{A_L}Y^\dagger_{B_L}),
 \ee
where
 \be
 Y^A=(A_1,A_2,B^\dagger_1, B^\dagger_2),\hspace{3ex}
 Y^\dagger_A=(A_1^\dagger,A_2^\dagger,B_1, B_2).
 \ee
In general, the leading order ADM of this class of composite
operators can be identified with the Hamiltonian of an $SU(4)$
spin chain with sites alternating between the fundamental and
anti-fundamental representations. We will not review the relevant
discussions here. There are three sets of Bethe roots, satisfying
the coupled Bethe equations.

Let us consider $SU(2)\times SU(2)$ subsector of the $SU(4)$ spin
chain, in which $Y^A$ take only $A_1,A_2$ and $Y^\dagger_A$ take
only $B_1,B_2$. Obviously the operator dual to the circular string
and the folded string ($\omega_1+\omega_2>0$ case) belong to this
subsector. In this case, we have two decoupled $SU(2)$ chains. The
middle Bethe roots $r_j$ in \cite{Minahan:2008hf} will not appear in
this case. And the Bethe equations are: \be
\left(\frac{u_j+i/2}{u_j-i/2}\right)^L=\prod_{k=1, k\ne
j}^{M_u}\frac{u_j-u_k+i}{u_j-u_k-i}, \ee \be
\left(\frac{v_j+i/2}{v_j-i/2}\right)^L=\prod_{k=1, k\ne
j}^{M_v}\frac{v_j-v_k+i}{v_j-v_k-i}. \ee The trace condition is \be
1=\prod_{j=1}^{M_u}\frac{u_j+1/2}{u_j-1/2}\prod_{j=1}^{M_v}\frac{v_j+1/2}{v_j-1/2}.
\ee The energy is \be
E={\lambda^2}(\sum_{j=1}^{M_u}\frac1{u_j^2+\frac14}+\sum_{j=1}^{M_v}\frac1{v_j^2+\frac14}).\ee
Here we have $L=2J$ and $M_u=M_v=L/2$.


One choice to satisfy the trace condition is to satisfy \be
1=\prod_{j=1}^{M_u}\frac{u_j+1/2}{u_j-1/2},\ee and
\be1=\prod_{j=1}^{M_v}\frac{v_j+1/2}{v_j-1/2}, \ee separately. Then
these two chains are totally unrelated.

For one chain in the large $J$ limit, the Bethe roots may distribute in two different ways,
as discussed in \cite{spinning}. For the circular string, the Bethe
roots distribute along the imaginary axis. It is natural to expect
that for the circular string in this paper, the Bethe roots
distribution follow the same way. Then use the results in
\cite{spinning}, we get that in the large $N$ and large $J$ limit, the anomalous
dimension of this operator: \be \gamma=\frac{4\pi^2\lambda^2}J.\ee
So the dimension of this operator obtained from the field theory
side at the weak coupling is:
\bea\Delta&=&2J+\frac{4\pi^2\lambda^2}J+\cdots\nn\\&=&J_1(1+\frac{\tilde{\lambda}^2}{4J_1^2}+\cdots).
\eea Comparing this result with the string energy (\ref{circular}),
we find that at large $J$ limit, the energy is proportional to $J_1$
at the leading order and the first order correction is always
$1/J_1^2$. However, on string side the first order contribution is
linear in $\tilde{\lambda}$, while on the field theory side, it is
quadratic in $\tilde{\lambda}$. This is very similar to what happens
in the point-like string case, where the string and field theory
result on fluctuation spectrum has a mismatch of factor $\lambda$.

The folded string case is more impressive. Let us just consider the
dual operators
$\Tr((A_1B_1)^{\frac{J_1}{2}+J_2}(A_2B_2)^{\frac{J_1}{2}-J_2})$,
which belong to the decoupled subsector. For a single subsector, the
operator looks like the same one corresponding to the IIB folded
string in $AdS_5\times S^5$ with two angular momenta, namely the
operators of the form $\Tr Z^{J_1}\Phi^{J_2}+\cdots $. In IIB case,
the match between the string  and the field theory result is in a
highly nontrivial way\cite{BFST2003}. Without getting into details,
we can show that our folded string is also in good match with the
field theory result, up to an interpolating functin of the 't Hooft
coupling constant. Notice that the relation (\ref{relation}) is
the same as the relation (2.2) in \cite{BFST2003}, after identifying
the angular momenta properly. And since we have the similar
integrable structure, the discussion in \cite{BFST2003} could be
applied to our case straightforwardly \footnote{The only difference is that
now the first non-trivial correction in the field theory appears at the order
of $\lambda^2$, so an interpolating fuction of $\lambda$ is needed here. See also
our discussions in the next section.}.
In this way, we show that for
the folded string, we have very good first leading order match in
$AdS_4/CFT_3$ correspondence.

For other circular string and folded string, the study of the dual
operators is quite similar.  the dual operators
should belong to the decoupled  $SU(2)\times SU(2)$ subsector.
And the relation between energy and the
angular momenta in these cases is consisent with this statement.



\section{Conclusion}

In this paper, we studied the semi-classical string configurations
in $AdS_4\times CP^3$ and their possible field theory dual. We
constructed point-like, circular, folded and pulsating strings in
$CP^3$ and calculate their energy and angular momenta. For the
circular strings and one class of folded string configurations, we
figured out their field theory dual operators. For one class of
circular string, the dual operators fall into the $SU(2)\times
SU(2)$ subsector of an integrable $SU(4)$ spin chain. In this
subsector, the fact that the two spin chains decouple allow us to
calculate the eigenvalues of ADM from Bethe equations. On the field
side, the ADM get correction only at two-loop order, which is
proportional to $\lambda^2$. This suggest that in the large angular
momentum limit, an effective expansion parameter could be
$\lambda^2/J^2$. On the other hand, from the string calculation, we
learn that the first order correction is linear in the expansion
parameter  $\lambda/J^2$. As suggested in
\cite{Gaiotto:2008cg,Nishioka:2008gz,Grignani:2008is}, there should
exist a interpolating function $f(\lambda)$ which approaches
$\lambda$ at weak coupling and $\sqrt{\lambda}$ at strong coupling.
Combining the result we found in this paper, we suggest that in the
study of the spinning strings, the effective expansion parameter is
$f^2(\lambda)/J^2$. The nontrivial functional match in the first
leading order in the folded string case gives very strong support to
this suggestion.



\section*{Acknowledgments}
The work was partially supported by NSFC Grant No.10535060,10775002
and NKBRPC (No. 2006CB805905). JB would like to thank Matteo
Bertolini, Bo Feng, Edi Gava, Chethan Gowdigere, Mahdi Torabian and
Ho-Ung Yee for helpful discussions.

  \end{document}